\documentclass[twocolumn,english,showpacs,prl,aps,superscriptaddress,floatfix]{revtex4}
\usepackage[T1]{fontenc}
\usepackage[latin9]{inputenc}
\usepackage{float}
\usepackage{amsmath}
\usepackage{amssymb}
\usepackage{graphicx}
\usepackage{esint}

\makeatletter

\providecommand{\tabularnewline}{\\}

\@ifundefined{textcolor}{}
{%
 \definecolor{BLACK}{gray}{0}
 \definecolor{WHITE}{gray}{1}
 \definecolor{RED}{rgb}{1,0,0}
 \definecolor{GREEN}{rgb}{0,1,0}
 \definecolor{BLUE}{rgb}{0,0,1}
 \definecolor{CYAN}{cmyk}{1,0,0,0}
 \definecolor{MAGENTA}{cmyk}{0,1,0,0}
 \definecolor{YELLOW}{cmyk}{0,0,1,0}
 }

\usepackage{float}\usepackage{amsfonts}\usepackage{amsthm}
\usepackage{bbm}

\usepackage{babel}

\makeatother

\begin{document}

\title{Merging of Dirac points and Floquet topological transitions in AC
driven graphene}

\author{Pierre Delplace}

\thanks{These authors contributed equally to this work.}

\affiliation{D\'epartement de Physique Th\'eorique, Universit\'e de Gen\`{e}ve, CH-1211
Gen\`{e}ve, Switzerland}

\author{\'Alvaro G\'omez-Le\'on}

\thanks{These authors contributed equally to this work.}

\author{Gloria Platero}

\affiliation{Instituto de Ciencia de Materiales, CSIC, Cantoblanco, Madrid E-28049,
Spain}

\date{\today}

\begin{abstract}
We investigate the effect of an in-plane AC electric field coupled
to electrons in the honeycomb lattice and show that it can be used
to manipulate the Dirac points of the electronic structure. We find
that the position of the Dirac points can be controlled by the amplitude
and the polarization of the field for high frequency drivings, providing
a new platform to achieve their merging, a topological transition
which has not been observed yet in electronic systems. Importantly,
for lower frequencies we find that the multi-photon absorptions and
emissions processes yield the creation of additional pairs of Dirac
points. This provides an additional method to achieve the merging
transition by just tuning the frequency of the driving. Our approach,
based on Floquet formalism, is neither restricted to specific choice
of amplitude or polarization of the field, nor to a low energy approximation for the Hamiltonian.
\end{abstract}
\maketitle

\section{Introduction}

Since the recent discovery of topological insulators\cite{hasan10,qi11},
the search for topological transitions in condensed matter has become
a priority task. In particular, the prediction and the observation
of a new topological order, namely the $\mathbb{Z}_{2}$ topological
order\cite{kane_z2_2005,bernevig06,konig07}, has stimulated a great
interest in the scientific community. Graphene, a two-dimensional
carbon-based crystal, is a wonderful platform for investigating topological
transitions. It is a semi-metal, whose band structure of the $p_{z}$
orbitals consists in two bands that touch linearly at the Fermi level
at two inequivalent points of the Brillouin zone\cite{castrorevue09}.
This peculiar structure gives rise to massless Dirac-like low energy
excitations.

On the other hand, several ideas to induce topological states of matter
by means of time periodic external potentials have been proposed\cite{inoue10,linder11,gu11,Kitagawa_Graphene,AGL_GP_Floquet,cayssol},
and recently observed in temporal modulated photonic crystals \cite{rechtsman13}.
This approach renews considerably the possibilities of inducing new
topological phases. Remarkably in graphene, a different topological
transition between a Dirac semi-metallic phase and an insulating phase,
can also be achieved by merging the pair of Dirac points (PDP)\cite{dietl08,guinea08,pereira_strain}.
The resulting insulating phase is not a $\mathbb{Z}_{2}$ topological
phase, but may nevertheless host zero-energy edge states whose topological
origin is well understood in terms of a one-dimensional bulk winding
number, namely the Zak phase\cite{zak,hatsugai02,delplace_zak}. The
search for this transition has stimulated experimental efforts beyond
the solid state community: The merging of the Dirac points, as
well as the emergence of zero-energy edge states have been recently
observed in anisotropic traps of cold atoms\cite{tarruell12,lim12}
and in microwave tight-binding analogue experiments of a honeycomb
lattice\cite{rechtsman12,mortessagne13}. However, the merging transition
has not been observed yet in electronic systems. In particular its
achievement in graphene itself\cite{montambaux09}, by means of mechanical
manipulations like stretching, is unfortunately out of reach, because
the graphene sheet would be destroyed far before the expected transition\cite{pereira_strain}.

\begin{figure}[H]
\centering{}\includegraphics[scale=0.22]{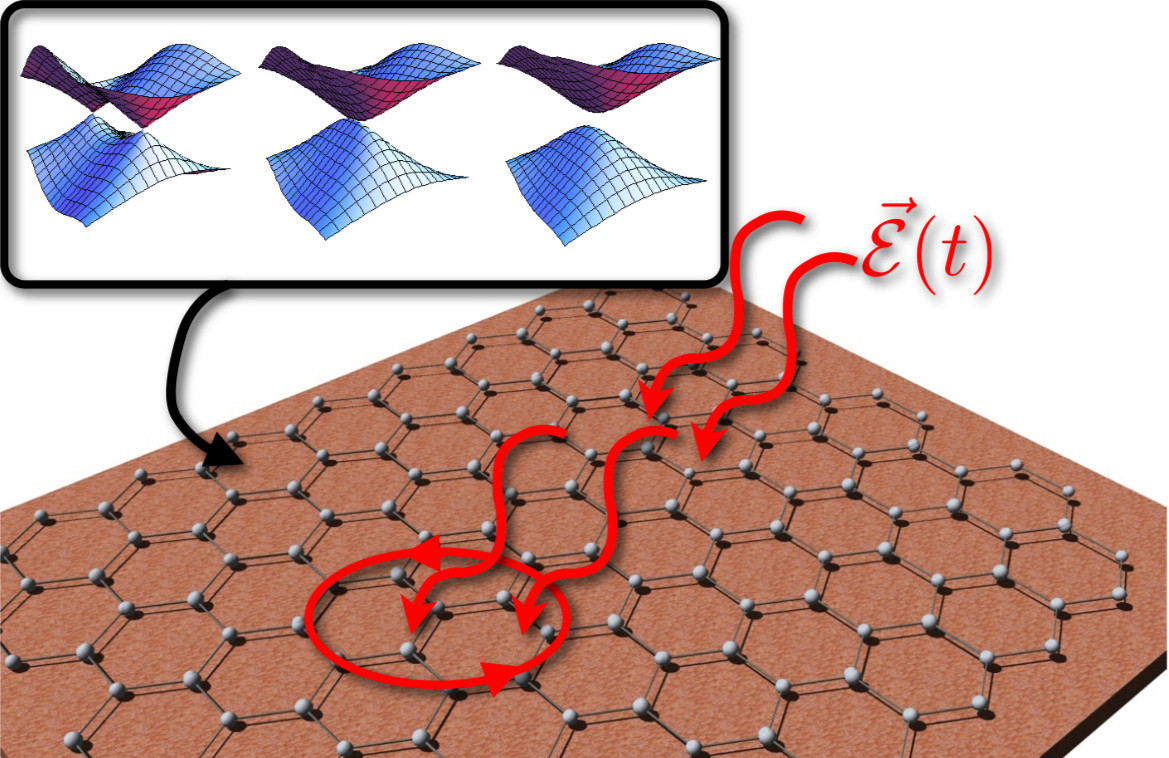} \caption{\label{fig:Schematic}Honeycomb lattice irradiated by an AC electric field
with arbitrary polarization. The inset shows the merging of the Dirac
points induced by the external field.}
\end{figure}

In the present work, we propose an alternative to mechanical distortions
to achieve the merging transition in the honeycomb lattice. In particular, 
we show that an AC electric field can drive the topological merging
 transition, or inversely induce the creation of new Dirac points.
This mechanism is different to the ones usually studied in previous
works dealing with Floquet topological insulators and summarized in
\cite{cayssol}. More precisely, we demonstrate that in the high frequency
regime, the AC field acts similarly to a mechanical strain, and therefore
allows for the manipulation of the Dirac points, their annihilation
and their creation in a controllable way. Furthermore, the AC field
is able to induce the localization of the electrons in specific directions
by aligning the Dirac points, which marks another topological transition
between two insulating or two semi-metallic phases. Our analysis,
when restricted to linear polarization and high frequency regime,
agrees with the results obtained in shaken optical lattices\cite{koghee12}.
In addition, we also consider different field polarizations and finite
frequency effects. Importantly, at lower frequencies, we find that
the coupling between the Floquet bands, that characterize the properties
of the driven system, gives rise to out-of-equilibrium phases with
no analogue in the static system in which multiple PDP emerge. Our approach
does not rely neither on low energy models nor on approximations valid
just close to resonance (as the rotating wave approximation)\cite{Maria,gu11,linder11},
but takes into account the full lattice model together with arbitrary field 
amplitudes and phase polarizations. This allows us to keep track of
the relative position between the Dirac points and thus properly describe
their annihilation or creation.

\section{Floquet theory on the honeycomb lattice}

The Hamiltonian of a $T$-periodic driven system fulfills $H\left(\tau+T\right)=H\left(\tau\right)$,
and its eigenvectors can then be written in the Floquet form: $|\psi\left(\tau\right)\rangle=e^{-i\epsilon\tau}|\Phi\left(\tau\right)\rangle$
($\hbar=1$), being $\epsilon$ the quasi-energy, and $|\Phi\left(\tau\right)\rangle=|\Phi\left(\tau+T\right)\rangle$
the Floquet state\cite{gloria04,AGL_GP_Floquet}. This ansatz, consequence
of the time translation invariance, maps the time-dependent Schr\"odinger
equation to the eigenvalue equation $\mathcal{H}\left(t\right)|\Phi\left(\tau\right)\rangle=\epsilon|\Phi\left(\tau\right)\rangle$,
in which the quasi-energies and the Floquet states are the eigenvalues
and eigenvectors of the Floquet operator $\mathcal{H}\left(\tau\right)\equiv H\left(\tau\right)-i\partial_{\tau}$,
respectively. In order to deal with the time dependence of the Floquet
operator $\mathcal{H}\left(\tau\right)$, it is useful to introduce
the Sambe space, which consists in a composed Hilbert space, where
the basis states are time-independent\cite{sambe73}. In this space,
the quasi-energies are given by (see \cite{AGL_GP_Floquet} and Appendix
A): 
\begin{equation}
\epsilon=\sum_{p,p'}\langle\Phi_{p'}|H_{p'-p}|\Phi_{p}\rangle-\ \delta_{p,p'}p\omega\label{eq:epsilon}
\end{equation}
where $H_{p'-p}=\int_{0}^{T}\frac{d\tau}{T}e^{i\omega\tau(p'-p)}H(\tau)$
with $\omega=2\pi/T$, and where $\Phi_{p^{(\prime)}}$ is the $p^{(\prime)}-\text{th}\in\mathbb{Z}$
Fourier component of the Floquet state. The coupling between the Floquet
side-bands is encoded in the Fourier components $H_{p'-p}$ of the
Hamiltonian.

For a honeycomb lattice embedded in a periodic time-dependent in-plane
electric field $\mathbf{{\cal E}}(\tau)$ (see Fig.\ref{fig:Schematic}),
the Peierls substitution leads to a time dependent Hamiltonian: 
\begin{equation}
H(\tau,\mathbf{k})=\left(\begin{array}{cc}
0 & \rho(\tau,\mathbf{k})\\
\rho^{*}(\tau,\mathbf{k}) & 0
\end{array}\right),\label{eq:ham}
\end{equation}
where $\rho(\tau,\mathbf{k})=\sum_{j}t_{j}(\tau)e^{i\mathbf{k}\cdot\mathbf{a}_{j}}$,
$t_{j}(\tau)=te^{i\mathbf{d}_{j}\cdot\mathbf{A}(\tau)}$ being $t$ the nearest neighbor hopping parameter, $\mathbf{a}_{j=1,2}$
are the basis vectors of the Bravais lattice: $\mathbf{a}_{1}=\frac{a}{2}\left(3,-\sqrt{3}\right)$,
$\mathbf{a}_{2}=\frac{a}{2}\left(3,\sqrt{3}\right)$, and $\mathbf{a}_{3}=a\left(0,0\right)$.
The vectors joining the nearest neighbors $\mathbf{d}_{j}$ are given
by $\mathbf{d}_{1}=\frac{a}{2}\left(-1,\sqrt{3}\right)$, $\mathbf{d}_{2}=\frac{a}{2}\left(-1,-\sqrt{3}\right)$,
$\mathbf{d}_{3}=a\left(1,0\right)$. We consider electric fields with
arbitrary amplitude and polarization, and write the vector potential
as $\mathbf{A}(\tau)=(A_{x}\sin(\omega\tau),A_{y}\sin(\omega\tau+\varphi))$.
The calculation of the Fourier components $H_{q=p'-p}(\mathbf{k})$
can be performed analytically (see Appendix B) and yields: 
\begin{equation}
H_{q}(\mathbf{k})=\left(\begin{array}{cc}
0 & \rho_{q}(\mathbf{k})\\
\rho_{-q}^{*}(\mathbf{k}) & 0
\end{array}\right),\label{eq:floq}
\end{equation}
with $\rho_{q}(\mathbf{k})=\sum_{j}t_{j,q}^{F}e^{i\mathbf{k}\cdot\mathbf{a}_{j}}$,
and where we have defined the time independent Floquet hoppings $t_{j,q}^{F}=tJ_{-q}({\cal A}_{j})e^{iq\Psi_{j}}$,
being $J_{q}\left(x\right)$ the $q^{\text{th}}$ order Bessel function
of the first kind. The dimensionless functions ${\cal A}_{j}$ and
$\Psi_{j}$ encode all the information of the field configuration:
\begin{equation}
\begin{split}{\cal A}_{2,1}= & \frac{a}{2}\sqrt{A_{x}^{2}+3A_{y}^{2}\pm2\sqrt{3}A_{x}A_{y}\cos\left(\varphi\right)},\\
\Psi_{2,1}= & \pm\arctan\left[\frac{\sqrt{3}A_{y}\sin\left(\varphi\right)}{A_{x}\pm\sqrt{3}A_{y}\cos\left(\varphi\right)}\right],\\
{\cal A}_{3}= & A_{x}a,\quad\Psi_{3}=0\ .
\end{split}
\label{eq:functions}
\end{equation}
 As a consequence, a spatial anisotropy can be tuned by varying the
polarization or the amplitude of the field. This is the key result
of the present work.

For the sake of clarity, we shall first focus on the high frequency
regime $\omega\gg t$. In that limit, the Floquet bands are uncoupled
and the zeroth Fourier component ${\cal H}_{q=0}$ of the Floquet
operator is the dominant one. Thus, Eq.\ref{eq:epsilon} is block-diagonal
in the Fourier space, and simply consists in a collection of identical
$2\times2$ time-independent Hamiltonians separated in energy by $\omega$.
The quasi-energy of a side-band $\alpha$, is then simply given by
$\epsilon_{\alpha}(\mathbf{k};A_{x},A_{y},\varphi)=\pm|\rho_{q=0}(\mathbf{k};A_{x},A_{y},\varphi)|+\alpha\omega$,
which shows that the quasi-energy of each Floquet band is, in the
high-frequency regime, identical to the energy bands of the undriven
honeycomb lattice with renormalized hopping parameters $t_{j,0}^{F}(A_{x},A_{y},\varphi)=tJ_{0}({\cal A}_{j})$.

\section{Manipulation of the Dirac points by the AC field: merging and localization}

In this section, we analyze the fate of the Dirac points of the quasi-energy
spectrum in the high frequency regime as a function of the parameters
of the electric field. Note that in this regime, the renormalized
hoppings $tJ_{0}({\cal A}_{j})$ are real parameters, so that the
system is still time-reversal invariant for all field polarizations,
and the Dirac nodes are thus well defined\cite{GDP-AQHE}. A direct
consequence of the AC field induced anisotropy of the hopping parameters,
is to change the location of the Dirac points. The two Dirac points,
related by time-reversal and inversion symmetry, move as the anisotropy
is modified and merge at one of the four time-reversal (inversion)
symmetric points of the Brillouin zone M$_{i}$, whenever a specific
relation between the hopping parameters is fulfilled \cite{guinea08,montambaux09,merginguniv09}:
\begin{equation}
\begin{split} & \text{M}_{0},\quad t_{1}^{F}+t_{2}^{F}+t_{3}^{F}=0\qquad\ \text{M}_{1},\quad t_{1}^{F}=t_{2}^{F}+t_{3}^{F}\\
 & \text{M}_{2},\quad t_{2}^{F}=t_{1}^{F}+t_{3}^{F}\qquad\qquad\text{M}_{3},\quad t_{3}^{F}=t_{1}^{F}+t_{2}^{F}
\end{split}
\label{eq:condition}
\end{equation}
where the index $0$ has been dropped out for clarity. The merging
transition corresponds to the creation/annihilation of a PDP. At the
transition, the so-called semi-Dirac dispersion relation is quadratic
in one direction, but remains linear in the other one, leading to
the prediction of striking properties such as an unusual temperature
dependence of the specific heat and magnetic field dependence of the
Landau levels \cite{merginguniv09}. Such transitions can now be achieved
for specific values of amplitude $(A_{x},A_{y})$ and polarization
$\varphi$ of the electric field.

\begin{figure}[ht!]
\centering{}\includegraphics[scale=0.28,angle=-90]{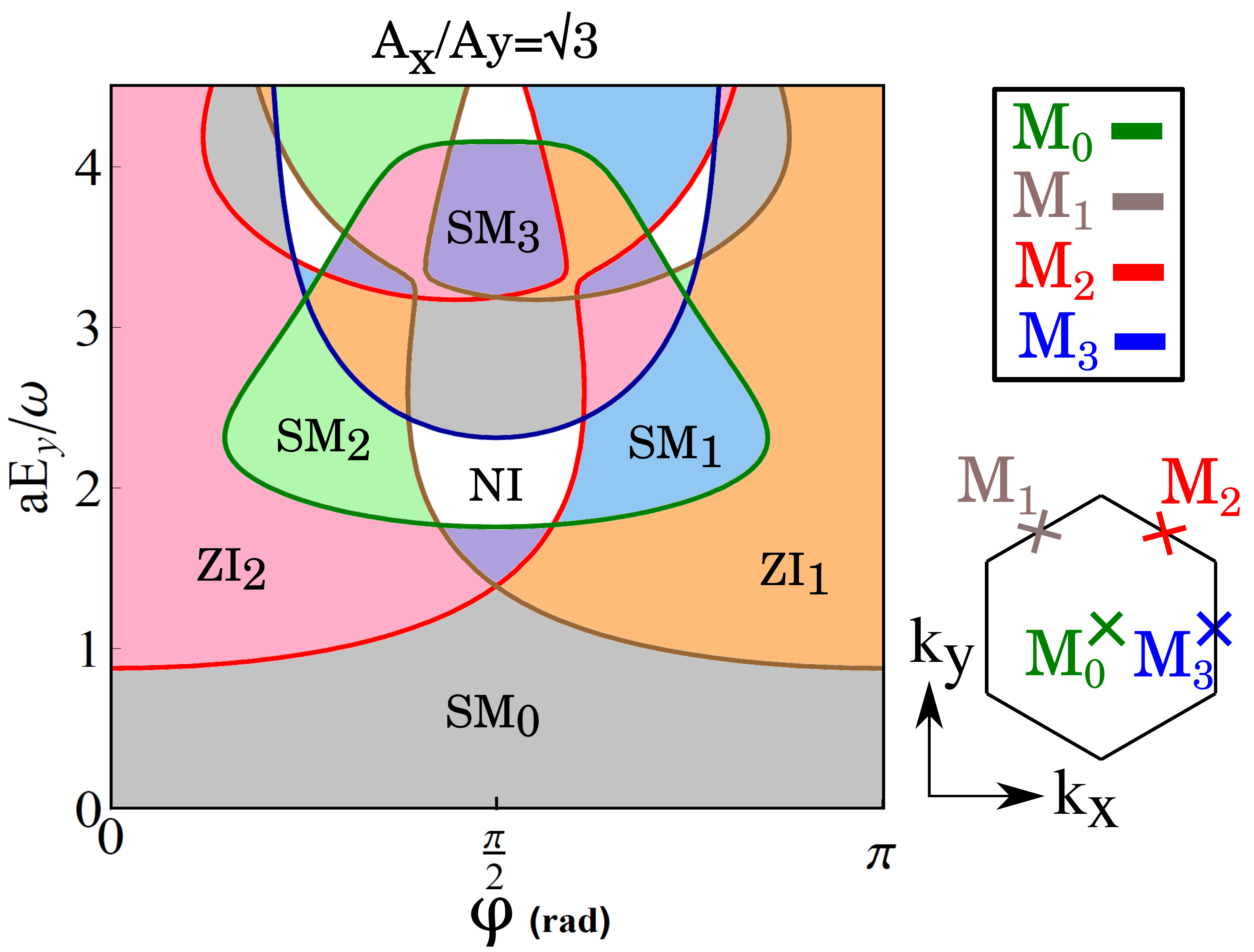} \caption{\label{fig:ellips}Semi-metallic (SM) / insulating (I) phase diagram
($\varphi,aE_{y}/\omega$) in the high frequency regime for fixed
$A_{x}/A_{y}=\sqrt{3}$, with $a$ the lattice spacing. NI (ZI) denotes
a normal (Zak) insulating phase. The different phases, which are 
represented in different colors, are separated by four distinct merging lines corresponding to the four M$_{i}$ points of the first Brillouin zone 
(shown on the right) where a PDP can be created/annihilated.}
\end{figure}

The system exhibits several distinct semi-metallic and insulating
phases that we now describe. Fig.\ref{fig:ellips} shows a typical
example of a semi-metallic/insulating phase diagram obtained in the
high frequency regime when varying the amplitude and the polarization
of the field (other examples are shown in the Appendix C). The creation/annihilation
of a PDP at a M$_{i}$ point is represented by a continuous \textit{merging
line} that separates a semi-metallic phase from an insulating phase.
Thus, there are four different merging lines, one for each point M$_{i}$.
It follows that distinct semi-metallic or insulating phases (coloured differently) may emerge when different merging lines are crossed. This distinction can be
made in terms of both the points M$_{i}$ and topological (winding)
numbers: Each semi-metallic phase can be labeled by a point M$_{i}$
where a PDP \textit{cannot} be annihilated. For instance, a PDP cannot
be annihilated at M$_{0}$ (green line) from the semi-metallic phase
$\text{SM}_{0}$ (which is the one connected to the undriven graphene
phase). The reason is that all the hopping parameters in the $\text{SM}_{0}$
phase have the same sign, so that the merging at $\text{M}_{0}$ would
require first a change in sign of any of them, and according to Eq.\ref{eq:condition},
it would fulfill the merging conditions at other M$_{i\neq0}$ point.
Therefore, one can distinguish four semi-metallic phases, denoted
as $\text{SM}_{i}$, one for each $\text{M}_{i}$ point where the
merging transition cannot occur. In addition, two winding numbers
can be introduced to characterize the topology of these four semi-metallic
phases. Indeed, the topological properties of out-of-equilibrium Floquet
systems can be characterized in a similar way to those of static systems
(see for instance Ref.\cite{AGL_GP_Floquet}). The first one is the
quantized Berry phase around a Dirac point, giving rise to a topological
charge $\frac{1}{2\pi}\oint d\mathbf{k}\cdot\nabla_{\mathbf{k}}\theta_{\mathbf{k}}=\pm1$
assigned to each Dirac point, where $\theta_{\mathbf{k}}=\text{arg}\rho(\mathbf{k})$
is the polar angle parameterizing the Bloch sphere. The two Dirac
points of graphene carry opposite topological charges that annihilate
at the merging transition \cite{montambaux09}. Besides, the topological
transition is accompanied with a change of a second winding number,
\textit{the Berry phase evaluated across a reduced one-dimensional
Brillouin zone}, namely the Zak phase $Z=\frac{1}{2\pi}\int_{-\mathbf{G}/2}^{\mathbf{G}/2}\left\{ d\mathbf{k}\cdot\nabla_{\mathbf{k}}\theta_{\mathbf{k}}\right\} $,
where $\mathbf{G}$ is a vector of the reciprocal lattice. Chiral
symmetry guaranties integer values of the Zak phase \cite{hatsugai02}
which depend on the direction in the reciprocal space. This reflects
the edge orientations dependence for the density of zero-energy edge
modes \cite{delplace_zak}. Thus, the four semi-metallic phases differ
by the set of values the Zak phase takes for all directions in the
reciprocal space, and therefore by the range of existence in $k$-space
of edge zero-energy modes for a given edge orientation. 
Similarly, different insulating phases can be distinguished as well
by the set of values the Zak phase takes in all possible directions,
even though a PDP can, in principle, be created at any of the four
points M$_{i}$. We find one normal insulating (NI) phase for which
the Zak phase is zero in every directions (meaning the absence of
zero-energy edge states), and, for polarizations different than $\varphi=\pi/2$,
two \textit{Zak insulating} phases for which the Zak phase is non
zero in different directions. The Zak insulating phases reflect, in
2D, the underlying non-trivial topology expected for a BDI symmetry
class in 1D\cite{schnyder_classification_2009}.

\begin{figure}[ht]
\centering{}\includegraphics[width=0.43\textwidth]{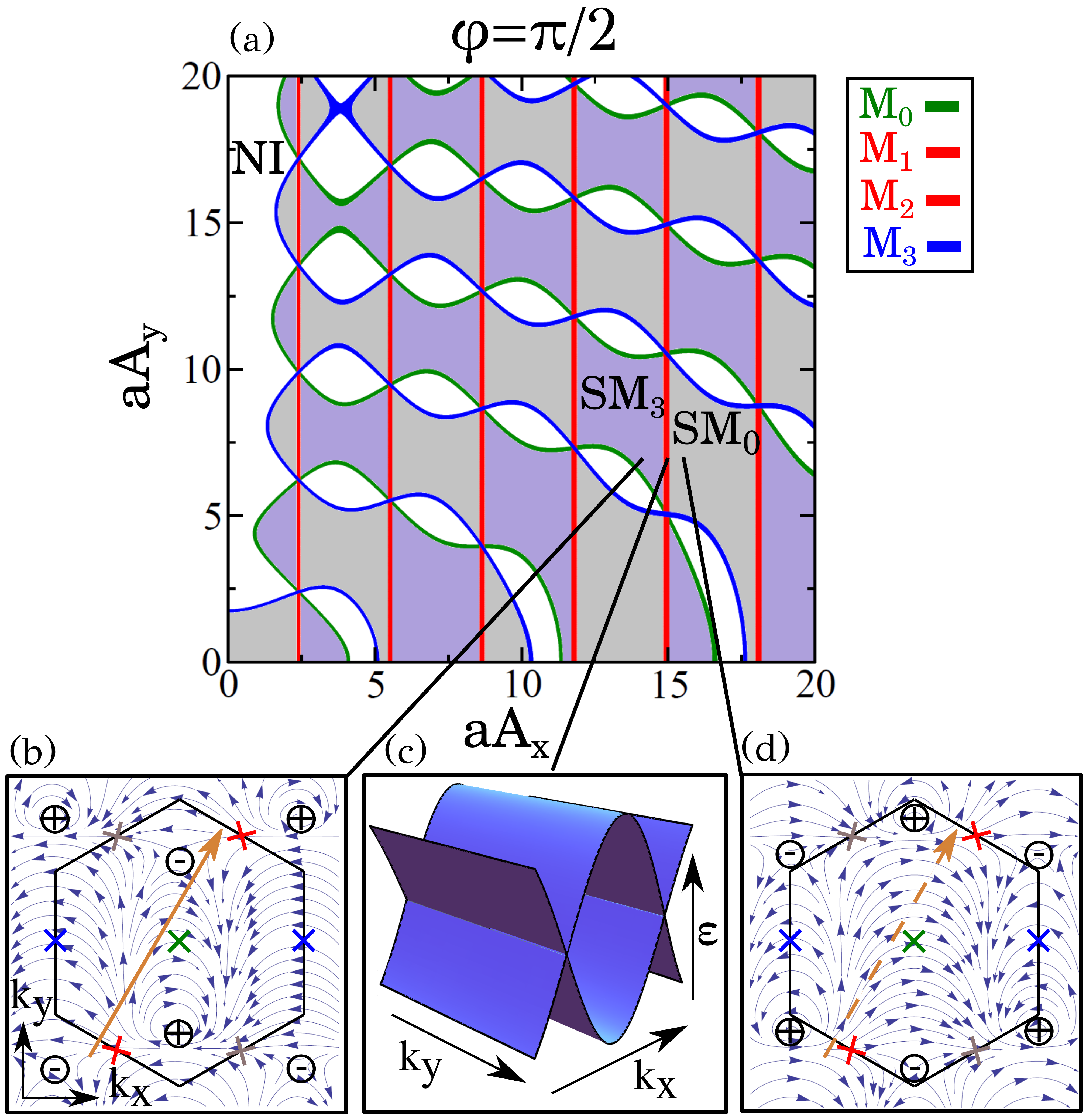} \caption{\label{fig:circular} (a) Semi-metallic (SM) / insulating (I) phase
diagram in the high frequency regime for $\varphi=\pi/2$, where only
three phases appear, due to the overlap between the M$_{1}$ and M$_{2}$
merging lines. (c) This overlap gives rise to localization lines separating
two semi-metallic phases. At the transition between these two semi-metallic
phases, the topological charges assigned to each Dirac point change
sign ((b) and (d)). This topological transition goes together with
a change of the value of the Zak phase, which is $1$ ($0$) in the
SM$_{3}$ (SM$_{0}$) phase along the path represented by a full (dashed)
arrow.}
\end{figure}

Interestingly, the phase diagrams show multiple crossings between
the merging lines, meaning that a PDP \textit{can be created and annihilated
simultaneously at two different $\text{M}_{i}$ points}. Following
Eq.\ref{eq:condition}, such crossings actually imply the vanishing
of one of the three hopping parameters $t_{j}^{F}$, what directly
induces localization of the electrons in the direction $\mathbf{d}_{j}$.
This striking feature corresponds to a transition between two insulating
or two semi-metallic phases. At the transition, the gap closes along
lines parallel to the $\mathbf{d}_{j}$ direction that passes through
the two distinct merging points. Unlike a single merging transition,
it follows that the dispersion relation is flat along the $\mathbf{d}_{j}$
direction but remains linear in the other one, as shown in Fig.\ref{fig:circular}
(c). At the transition, the system is then reduced to an array of
uncoupled one-dimensional chains, hosting one-dimensional massless
Dirac fermions. Remarkably, for phase polarization $\varphi=\pi/2$,
the merging at $M_{1}$ and $M_{2}$ is always coincident. This gives
rise to critical \textit{localization lines} that separate two SM
phases as shown in Fig.\ref{fig:circular}. At the transition between
two semi-metallic phases, the topological charges assigned to the
Dirac points change sign (see Fig.\ref{fig:circular} (b) and (d)).
In that sense, the localization (or double merging) transition is
also a topological transition. Finally, we report that the phase diagram
for polarization $\varphi=\pi/2$ also exhibits crossings between
the four merging lines simultaneously%
\footnote{(According to Eq. \ref{eq:condition}, the crossing of three merging
lines only is not possible.)%
}. At such critical points, all the hopping parameters $t_{j}^{F}$
vanish and the quasi-energy of each side-band is perfectly flat which
leads to full charge localization in all directions.

\section{Multi side-bands effects and emergence of additional Dirac points}

Although the merging transition in the high frequency regime is simple
to describe, it might be difficult to achieve experimentally in graphene,
since it requires high field amplitudes which also would involve heating effects. 
For that reason, we now investigate the effect of a frequency decrease and show 
that the merging transition may also be achieved by varying the frequency, and thus requires smaller intensities of the AC field. 
Importantly, besides being more relevant experimentally, it turns out that such a regime
is particularly interesting since it allows us to engineer additional Dirac points. 
\begin{figure}[ht!]
\centering{}\includegraphics[scale=0.26]{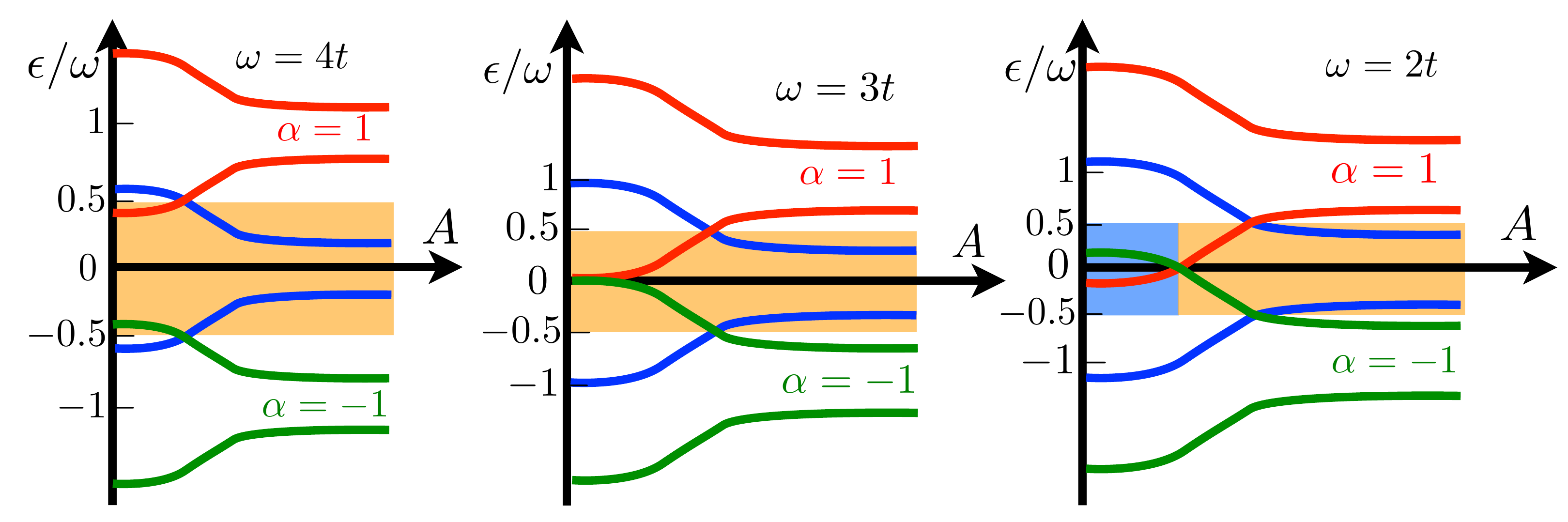} \caption{\label{fig:sketch} Sketch of the first uncoupled side-bands $\alpha=0,\pm1$
as a function of a parameter (for instance the field amplitude), for
different driving frequencies. The colored area highlight the range
where the side-bands are inverted.}
\end{figure}

The high frequency regime holds as long as the frequency is larger
than the bandwidth of the undriven system, namely, $\omega>6t$. For
lower frequencies ($3t<\omega\lesssim6t$) the coupling between the
Floquet side-bands becomes relevant, and the high frequency effective
theory is not accurate. Besides, unless the polarization of the field
is linear, the couplings break time reversal symmetry and will therefore
open a gap at the Dirac points%
\footnote{According to Eq.\ref{eq:functions}, non linear field polarizations
induce complex hoppings, and thus break time reversal symmetry opening
gaps. However, similarly to the Boron Nitride model, the topological
charges assigned to the non-trivial Berry phases in each valley remain,
so that their annihilation/creation can still be achieved by varying
the parameters of the field.%
}. In the following, we therefore restrict our analysis to linearly
polarized fields only, although our formalism remains valid for any
phase polarization. 
\begin{figure}[ht!]
\centering{}\includegraphics[scale=0.17]{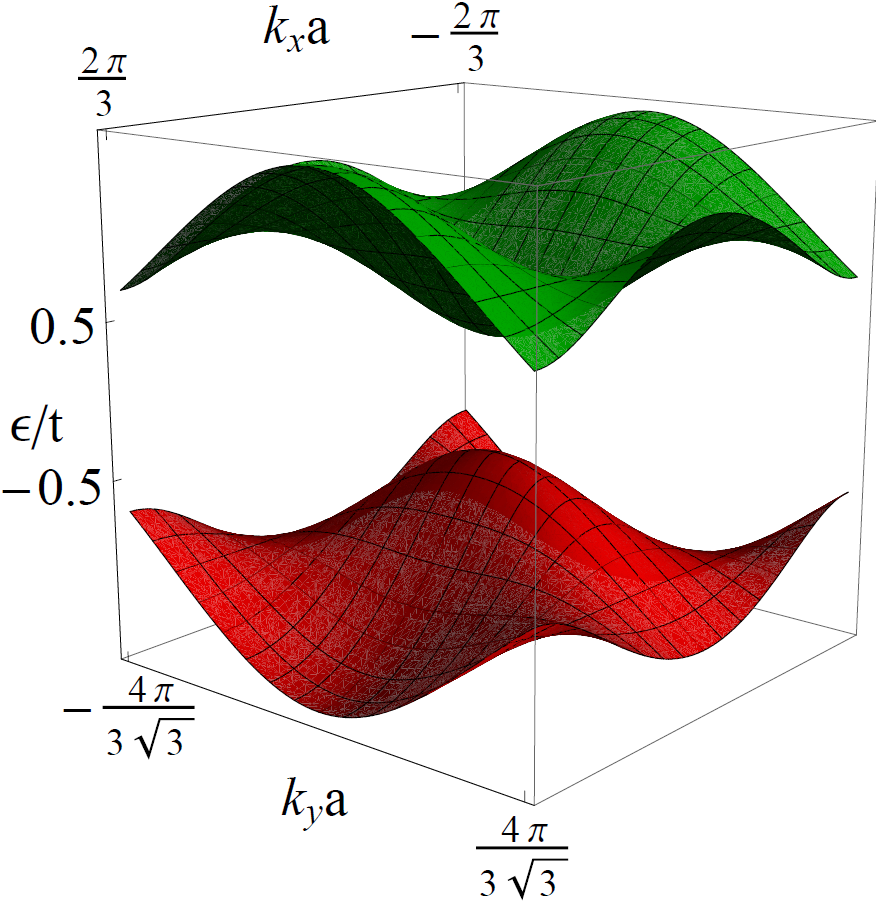}\includegraphics[scale=0.17]{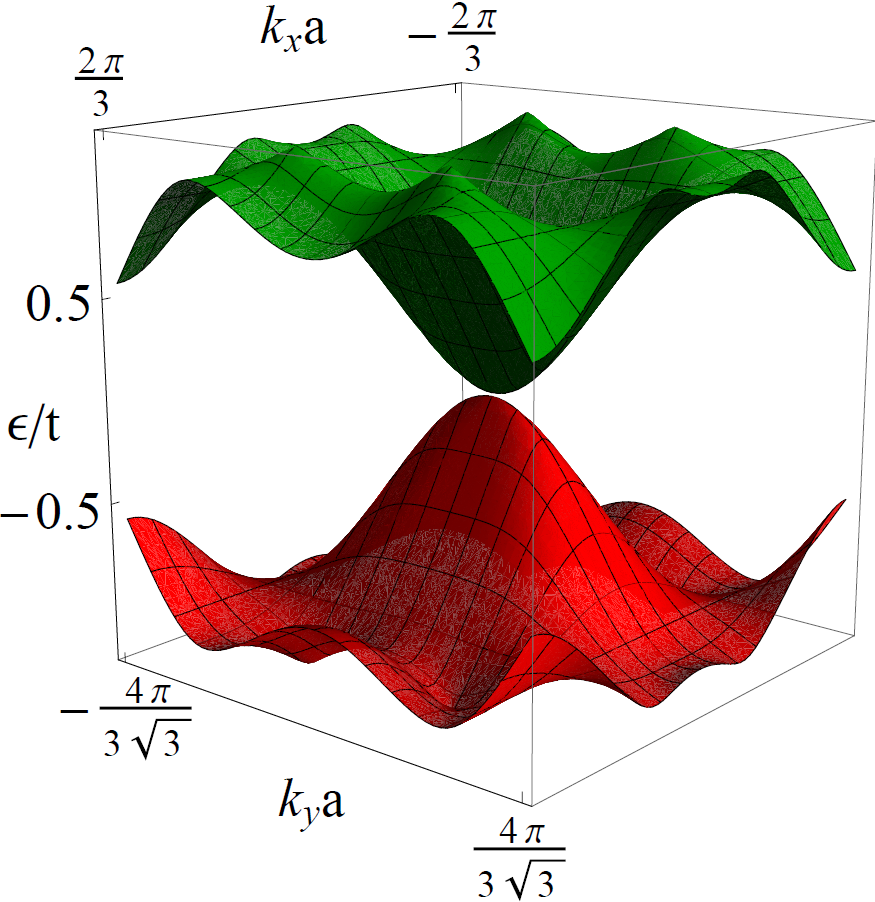}
\includegraphics[scale=0.19]{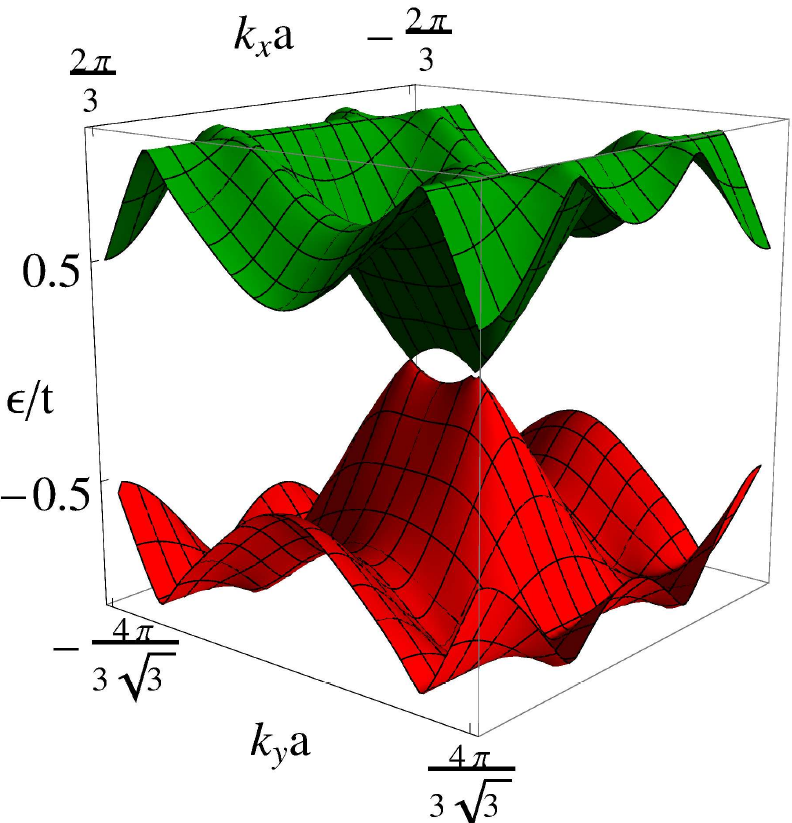}\includegraphics[scale=0.19]{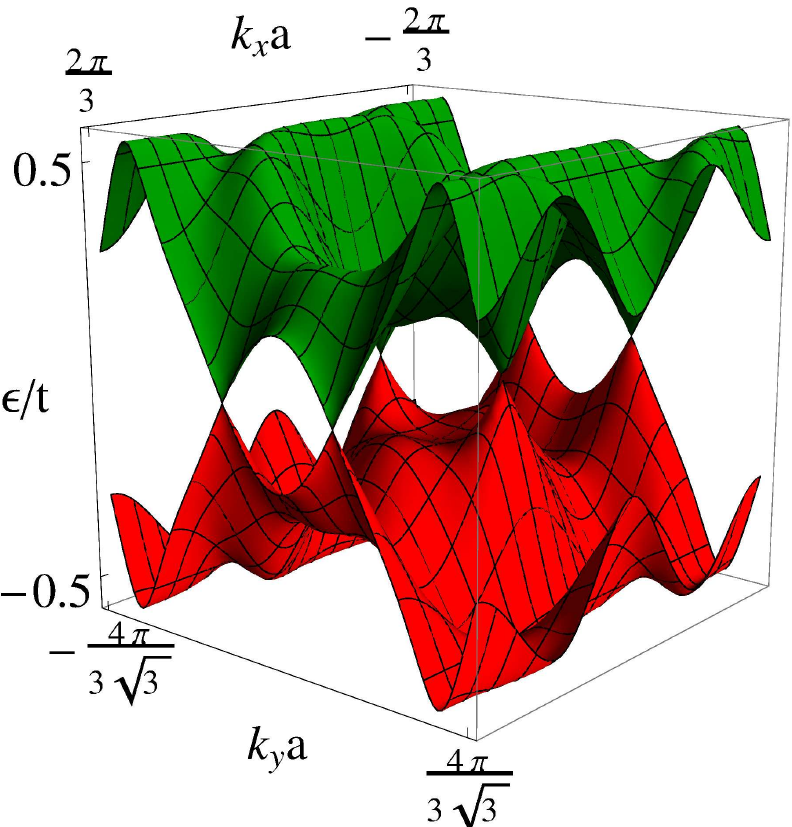}
\caption{\label{fig:low} Quasi-energy spectra for $A_{y}=3.6$, $A_{x}=0$
and $\varphi=0$ with (top, left) $\omega=10t$, (top, right) $\omega=2.5t$, (bottom, left) $\omega=2.1t$
and (bottom, right) $\omega=1.5t$. }
\end{figure}

As illustrated in Fig.\ref{fig:sketch}, the Floquet side-bands overlap
as the frequency is decreased. This yields new couplings which can
now include the absorption or the emission of photons. At around about
$\omega=3t$, the (uncoupled) side-bands $\alpha=+1$ and $\alpha=-1$
are expected to touch at $\epsilon=0$ for vanishing amplitude of
the field, giving rise to a new crossing at the $\text{M}_{0}$ point.
This new crossing corresponds to a band inversion, in which the symmetries
of the conduction band and the valence band are exchanged. As the high
frequency approximation fails to describe this situation, the Floquet
operator is now diagonalized numerically.

For more concreteness, we fix the system in a normal insulating phase
in the high frequency regime by choosing suitable field amplitudes
(Fig.\ref{fig:low}, top, left). Then, we decrease the frequency such that
the side-bands $\alpha=+1$ and $\alpha=-1$ cross. This leads to
the appearance of a PDP at the center of the Brillouin zone $\text{M}_{0}$
for $\omega=2.5t$ with the typical semi-Dirac dispersion relation
(quadratic in one direction and linear in the other one, as it is
shown in Fig.\ref{fig:low}, top, right). When the frequency is further decreased,
the two Dirac points move away to each other while the rest of the
bands bend (Fig.\ref{fig:low}, bottom, left). A larger decrease of frequency
makes the bands close the gap once again at the $\text{M}_{3}$ point
of the FBZ. This creates a \textit{second} PDP with the same typical
semi-Dirac shape, which move away as the frequency is lowered, as
it is shown in Fig.\ref{fig:low}, bottom, right. Then, for lower frequencies
the band structure now owns \textit{four} Dirac points at the same
quasi-energy. Therefore, the driving field offers not only the opportunity
to manipulate the Dirac points, but also to create new ones and observe
the merging transition by just tuning the frequency of the AC field.

\section{Discussion}

We have shown that the merging of the Dirac points in the honeycomb
lattice can be driven in a controllable way by an AC electric field.
We have obtained rich phase diagrams in which distinct topological
semi-metallic and insulating phases emerge. At high frequency, $\omega\gg t$,
the behavior is easily described by an effective Floquet operator,
equivalent to the Hamiltonian of undriven graphene, but with renormalized
hoppings tuned by the field amplitude and the phase polarization.
The resulting phases reflect interesting topological features, and
the critical lines in which pairs of Dirac points are created and
annihilated simultaneously show unexpected localization properties
for which electrons behave relativistically in one direction, while
they are localized in the other one.

Importantly, we also proved that out of the high frequency regime
($\omega\gtrsim t$), and for linearly polarized fields, a change
in frequency can drive a novel transition with the emergence of new
pairs of Dirac points, although it does not necessarily correspond
to a transition between an insulating and a semi-metallic phase. This
new transition lies on multi-side-bands couplings, or in other words,
multi-photon assisted transitions. The resulting new semi-metallic
phase with four Dirac points can clearly be distinguished from the
one with two Dirac points by various transport measurements. For instance,
a Hall conductance measurement would provide a direct signature of
the two additional Dirac points, since each valley brings a contribution
$e^{2}/h(1/2+n)$ to the total transverse conductivity, where $n$
is the number of Landau levels. The conductivity measurements would
be accomplished by considering the Floquet sum rule obtained in Ref.\cite{steady},
due to the far from equilibrium situation of the setup at intermediate
frequencies $\omega\sim t$. We also notice that for samples of the size of or smaller than the
wave length ($\approx1$ micron), the effect of the magnetic field
of the irradiation can be ignored.

For the realization in graphene, the high frequency regime requires
field frequencies in the near ultraviolet and field amplitudes at
least of the order $\approx3.4\text{V}\textrm{\AA}^{-1}$ \cite{Experiment1,Experiment2,Experiment3}.
The energy corresponding to this electric field is about half the
ionization energy of carbon atoms. However, real graphene also possesses $s$-orbitals
which can be affected by such a high frequency field. Thus, in order
to achieve the high frequency regime, a frequency larger than the actual band width of the graphene bands
must be considered. In addition, for fields of both high frequency and high amplitude,
the heating of the sample will certainly be an issue because of the
dissipative processes due to phonons and electron-electron scattering. One way to partially avoid these issues 
can be the use of alternative platforms with similar properties, such as artificial 
graphene\cite{Artificial-graphene}. This would help in two different ways: The increase
of the lattice parameter allows one to consider lower frequencies and thus lower the field intensities, 
while in addition, the contribution of the $s$-orbitals would vanish. 
Finally, we expect that the results obtained in the lower frequency regime  
($\omega\gtrsim t$) to be more easily achievable in real graphene, if only  
because the undesirable heating of the sample will be reduced.
Furthermore, we stress that this regime allows one not only to manipulate  
the Dirac points but also to create new ones just by tuning the frequency.\\

\textbf{Acknowledgments:} The authors would like to thank Janos Asboth, Jian Li, Markus B\"uttiker,
Hector Ochoa, Alexey Kuzmenko and F. Koppens for inspiring discussions.
P. D. was supported by the European Marie Curie ITN NanoCTM. \'A. G\'omez-Le\'on
acknowledges JAE program. \'A. G. L. and G. P. acknowledge MAT 2011-24331
and ITN, grant 234970 (EU) for financial support.

\appendix
\begin{widetext}

\section{Appendix A: Floquet-Bloch theory}

Crystals coupled to in-plane AC electric fields have lattice and time
translation invariance: $H\left(\mathbf{x}+\mathbf{a}_{i},\tau+T\right)=H\left(\mathbf{x}+\mathbf{a}_{i},\tau\right)=H\left(\mathbf{x},\tau+T\right)$,
being $\mathbf{a}_{i}$ the lattice vectors and $T=2\pi/\omega$ a
period of the driving field. Thus, one can assume solutions in Floquet-Bloch
form $|\Psi_{\alpha,\mathbf{k}}\left(\mathbf{x},\tau\right)\rangle=e^{i\mathbf{k}\cdot\mathbf{x}-i\epsilon_{\alpha,\mathbf{k}}\tau}|u_{\alpha,\mathbf{k}}\left(\mathbf{x},\tau\right)\rangle$,
where $\epsilon_{\alpha,\mathbf{k}}$ is the quasi-energy for the
$\alpha$ Floquet state, and $\mathbf{k}$ the wave-vector. The Floquet-Bloch
states $|u_{\alpha,\mathbf{k}}\left(\mathbf{x},\tau\right)\rangle$
are periodic in both $\mathbf{x}$ and $\tau$, and fulfill the Floquet
eigenvalue equation: 
\begin{eqnarray}
\mathcal{H}\left(\tau,\mathbf{k}\right)|u_{\alpha,\mathbf{k}}\rangle & = & \epsilon_{\alpha,\mathbf{k}}|u_{\alpha,\mathbf{k}}\rangle,\label{eq:Floquet-Eq}\\
\mathcal{H}\left(\tau,\mathbf{k}\right) & \equiv & e^{-i\mathbf{k}\cdot\mathbf{x}}\left(H\left(\tau\right)-i\partial_{\tau}\right)e^{i\mathbf{k}\cdot\mathbf{x}}\nonumber \\
 & = & H_{\mathbf{k}}\left(\tau\right)-i\partial_{\tau}\label{eq:Floquet-operator}
\end{eqnarray}
where $\mathcal{H}\left(\mathbf{k},\tau\right)$ is the Floquet operator.
The time dependent Hamiltonian is obtained through the Peierls substitution
in the tight binding model, leading to the time dependent hoppings
$t_{j}^{\alpha,\beta}\left(\tau\right)=t_{j}^{\alpha,\beta}e^{i\mathbf{A}\left(\tau\right)\cdot\mathbf{d}_{j}}$,
and to the time dependent Hamiltonian:

\begin{equation}
H\left(\tau,\mathbf{k}\right)=\sum_{\alpha,\beta}\sum_{j}c\left(\tau\right)_{\alpha,\mathbf{k}}^{\dagger}c\left(\tau\right)_{\beta,\mathbf{k}}t_{j}^{\alpha,\beta}\left(\tau\right)e^{i\mathbf{k}\cdot\mathbf{a}_{j}},\label{eq:H-minimal-coupling}
\end{equation}
where we have included the sub-lattice indexes $\alpha$, $\beta$,
and the time dependent annihilation and creation operators of Floquet-Bloch
states. Due to its time periodicity, we expand in Fourier series $c\left(\tau\right)_{\alpha,\mathbf{k}}$
and $c\left(\tau\right)_{\alpha,\mathbf{k}}^{\dagger}$: 
\begin{align}
H\left(\tau,\mathbf{k}\right) & =\sum_{\alpha,\beta}\sum_{p,p^{\prime}}\sum_{j}t_{j}^{\alpha,\beta}\left(\tau\right)e^{i\omega\tau\left(p^{\prime}-p\right)}e^{i\mathbf{k}\cdot\mathbf{a}_{j}}c_{\alpha,\mathbf{k},p^{\prime}}^{\dagger}c_{\beta,\mathbf{k},p}\ .\label{eq:Floquet-tight-Fourier}
\end{align}
Eq.\ref{eq:Floquet-tight-Fourier} gives a description of the time
dependent Hamiltonian in terms of the time independent operators $\left\{ c_{\alpha,\mathbf{k},p},c_{\alpha,\mathbf{k},p}^{\dagger}\right\} $.
The calculation of the quasi-energies is easily performed in Sambe
space by considering the composed scalar product:
\begin{equation}
\langle\langle\ldots\rangle\rangle=\frac{1}{T}\int_{0}^{T}\langle\ldots\rangle d\tau\ ,
\end{equation}
where $\langle\ldots\rangle$ represents the usual scalar product
in the Hilbert space. This leads to a set of Fourier components for
the time dependent Hamiltonian given by:
\begin{equation}
H_{q}\left(\mathbf{k}\right)=\sum_{\alpha,\beta}\sum_{j}\frac{1}{T}\int_{0}^{T}t_{j}^{\alpha,\beta}\left(\tau\right)e^{i\omega q\tau}e^{i\mathbf{k}\cdot\mathbf{a}_{j}}d\tau\ ,
\end{equation}
where $q=p^{\prime}-p$. The quasi-energies are then obtained by direct
diagonalization of the infinite dimensional Floquet operator in Sambe
space:
\begin{equation}
\mathcal{H}\left(\mathbf{k}\right)=\sum_{q,p}H_{q}\left(\mathbf{k}\right)-p\omega\delta_{q,0}\mathcal{I}\ ,\label{eq:Floquet-tight-binding}
\end{equation}
where $\mathcal{I}$ is the identity matrix for the sub-lattice subspace.
Note that this representation of the tight binding Hamiltonian simply
describes a set of hoppings: 
\begin{gather}
\tilde{t}_{p^{\prime},p}^{\alpha,\beta}\equiv\frac{1}{T}\int_{0}^{T}\sum_{j=1}^{3}e^{i\omega\tau\left(p^{\prime}-p\right)}e^{i\mathbf{k}\cdot\mathbf{d}_{j}}t_{j}^{\alpha,\beta}\left(\tau\right)d\tau,\label{eq:hoppings-nm}
\end{gather}
which include the emision or absorption of $q$ photons. The last
term of Eq.\ref{eq:Floquet-tight-binding} $p\omega\delta_{q,0}\mathcal{I}$
is simply the Fourier space representation of the time derivative
operator $-i\partial_{t}$.

\section{Appendix B: Periodically driven honeycomb lattice}

The honeycomb lattice is made of a primitive unit cell with two inequivalent
atoms (A,B) and unit cell translation vectors $\mathbf{a}_{1}=\frac{a}{2}\left(3,-\sqrt{3}\right)$,
and $\mathbf{a}_{2}=\frac{a}{2}\left(3,\sqrt{3}\right)$, where $a$
is the distance between the atom A and and the atom B. Each A atom
has three nearest neighbors (B type) at positions $\mathbf{d}_{3}=a\left(1,0\right)$,
$\mathbf{d}_{2}=\frac{a}{2}\left(-1,-\sqrt{3}\right)$ and $\mathbf{d}_{1}=\frac{a}{2}\left(-1,\sqrt{3}\right)$.
In $\mathbf{k}$-space the energy spectrum shows two inequivalent
gapless points at $\mathbf{K}=\frac{2\pi}{3a}\left(1,\frac{1}{\sqrt{3}}\right)$
and $\mathbf{K}^{\prime}=\frac{2\pi}{3a}\left(1,-\frac{1}{\sqrt{3}}\right)$
called Dirac points. We also define the Time Reversal Invariant Momentum
(TRIM) points as: 
\begin{eqnarray}
\text{M}_{0} & = & \left(0,0\right),\label{eq:Merging-points}\\
\text{M}_{1} & = & \frac{\pi}{3a}\left(-1,\sqrt{3}\right),\nonumber \\
\text{M}_{2} & = & \frac{\pi}{3a}\left(1,\sqrt{3}\right),\nonumber \\
\text{M}_{3} & = & \frac{2\pi}{3a}\left(1,0\right).\nonumber 
\end{eqnarray}
In absence of driving, the tight binding Hamiltonian up to nearest
neighbors (NN) for the honeycomb lattice is: 
\begin{equation}
H\left(\mathbf{k}\right)=\left(\begin{array}{cc}
0 & \rho\left(\mathbf{k}\right)\\
\rho\left(\mathbf{k}\right)^{*} & 0
\end{array}\right),\ \rho\left(\mathbf{k}\right)=\sum_{i=j}^{3}t_{j}e^{i\mathbf{k}\cdot\mathbf{a}_{j}},\label{eq:Undriven-graphene}
\end{equation}
where $t_{j}$ is referred to the hopping along the $\mathbf{d}_{j}$
direction, and for simplicity we have included $\mathbf{a}_{3}=a\left(0,0\right)$
to take into account the hopping within the same unit cell. Explicitly
$\rho\left(\mathbf{k}\right)=\sum_{j=1}^{3}\rho_{j}\left(\mathbf{k}\right)$
consists in the the sum of the next three terms: 
\begin{eqnarray}
\rho_{1}\left(\mathbf{k}\right) & = & te^{i\frac{a}{2}\left(3k_{x}-\sqrt{3}k_{y}\right)},\nonumber \\
\rho_{2}\left(\mathbf{k}\right) & = & te^{i\frac{a}{2}\left(3k_{x}+\sqrt{3}k_{y}\right)},\nonumber \\
\rho_{3}\left(\mathbf{k}\right) & = & t.
\end{eqnarray}
The spectrum of $H\left(\mathbf{k}\right)$ has two bands with dispersion
relation given by: 
\begin{equation}
E\left(\mathbf{k}\right)_{\pm}=\pm|\rho\left(\mathbf{k}\right)|=\pm t\sqrt{3+f\left(\mathbf{k}\right)},\label{eq:Graphene-energy}
\end{equation}
 
\begin{equation}
f\left(\mathbf{k}\right)\equiv2\cos\left(a\sqrt{3}k_{y}\right)+2\cos\left(\frac{3a}{2}k_{x}-\frac{\sqrt{3}a}{2}k_{y}\right)+2\cos\left(\frac{3a}{2}k_{x}+\frac{\sqrt{3}a}{2}k_{y}\right).
\end{equation}
When the system is coupled to an AC electric field in the dipolar
approximation, one can follow Appendix A to obtain the time dependent
Hamiltonian and the expression for the Floquet operator in Fourier
space. In general, we assume without loss of generality that the vector
potential for an homogeneous in-plane electric field is given by:
\begin{equation}
\mathbf{A}\left(\tau\right)=\left(A_{x}\sin\left(\omega\tau\right),A_{y}\sin\left(\omega\tau+\varphi\right),0\right).\label{eq:graphene-ac-field}
\end{equation}
which represents the case of an elliptically polarized field in the
x-y plane. Note that $\varphi=0$ reduces Eq.\ref{eq:graphene-ac-field}
to the case linear polarization, and $A_{x}=A_{y}$ with $\varphi=\pi/2$
reduces Eq.\ref{eq:graphene-ac-field} to the case of circular polarization.
To obtain the quasi-energies, one needs to calculate the matrix elements
of the Floquet operator by means of Eq.\ref{eq:Floquet-tight-binding}.
The integrals are of the form: 
\begin{eqnarray}
\frac{1}{T}\int_{0}^{T}e^{i\omega\tau\left(p^{\prime}-p\right)}e^{i\alpha\sin\left(\omega\tau\right)}e^{i\beta\sin\left(\omega\tau+\varphi\right)}d\tau & = & e^{i\left(p-p^{\prime}\right)\arctan\left(\frac{\beta\sin\left(\varphi\right)}{\alpha+\beta\cos\left(\varphi\right)}\right)}J_{p^{\prime}-p}\left(\sqrt{\alpha^{2}+\beta^{2}+2\alpha\beta\cos\left(\varphi\right)}\right).
\end{eqnarray}
which has been solved using the identity: 
\begin{equation}
e^{-i\nu\Psi}J_{\nu}\left(\Gamma\right)=\sum_{m=-\infty}^{\infty}J_{\nu+m}\left(\alpha\right)J_{m}\left(\beta\right)e^{-i\varphi m},
\end{equation}
where $\Gamma=\sqrt{\alpha^{2}+\beta^{2}-2\alpha\beta\cos\left(\phi\right)}$,
and $\Psi$ is obtained from the relation: 
\begin{equation}
\begin{cases}
\Gamma\cos\left(\Psi\right)= & \alpha-\beta\cos\left(\varphi\right)\\
\Gamma\sin\left(\Psi\right)= & \beta\sin\left(\varphi\right)
\end{cases},
\end{equation}
 i.e., 
\begin{equation}
\Psi=\arctan\left(\frac{\beta\sin\left(\varphi\right)}{\alpha-\beta\cos\left(\varphi\right)}\right).
\end{equation}
The Fourier space representation of the time dependent tight binding
is thus given by blocks ($q=p^{\prime}-p$): 
\begin{equation}
\tilde{t}_{p^{\prime},p}^{\alpha,\beta}=H_{q}\left(\mathbf{k}\right)=\left(\begin{array}{cc}
0 & \rho_{q}\left(\mathbf{k}\right)\\
\rho_{-q}^{*}\left(\mathbf{k}\right) & 0
\end{array}\right),
\end{equation}
with $\rho_{q}(\mathbf{k})=\sum_{j}t_{j,q}^{F}e^{i\mathbf{k}\cdot\mathbf{a}_{j}}$
and the renormalized hoppings: 
\begin{eqnarray}
t_{1,q}^{F} & = & te^{-iq\Psi_{1}}J_{q}\left(\mathcal{A}_{1}\right),\nonumber \\
t_{2,q}^{F} & = & te^{iq\Psi_{2}}J_{q}\left(\mathcal{A}_{2}\right),\nonumber \\
t_{3,q}^{F} & = & tJ_{q}\left(\mathcal{A}_{3}\right),
\end{eqnarray}
The functions $\Psi_{j}$ and $\mathcal{A}_{j}$ encode all the information
of the AC field configuration: 
\begin{eqnarray}
\mathcal{A}_{2,1} & = & \frac{a}{2}\sqrt{A_{x}^{2}+3A_{y}^{2}\pm2\sqrt{3}A_{x}A_{y}\cos\left(\varphi\right)}\ ,\\
\Psi_{2,1} & = & \arctan\left(\frac{\sqrt{3}A_{y}\sin\left(\varphi\right)}{A_{x}\pm\sqrt{3}A_{y}\cos\left(\varphi\right)}\right),\ \mathcal{A}_{3}=A_{x}a,\ \Psi_{3}=0\ .\nonumber 
\end{eqnarray}
In the high frequency regime ($\omega>6t$), the Floquet bands are
decoupled ($\mathcal{H}_{n,m}$ is block diagonal), and the system
can be described by an effective time independent Floquet operator,
which is given by the 2 by 2 matrix: 
\begin{equation}
\mathcal{H}^{\left(0\right)}=t\left(\begin{array}{cc}
0 & e^{-i\mathbf{k}\cdot\mathbf{a}_{1}}J_{0}\left(\mathcal{A}_{1}\right)+e^{-i\mathbf{k}\cdot\mathbf{a}_{2}}J_{0}\left(\mathcal{A}_{2}\right)+J_{0}\left(\mathcal{A}_{3}\right)\\
e^{i\mathbf{k}\cdot\mathbf{a}_{1}}J_{0}\left(\mathcal{A}_{1}\right)+e^{i\mathbf{k}\cdot\mathbf{a}_{2}}J_{0}\left(\mathcal{A}_{2}\right)+J_{0}\left(\mathcal{A}_{3}\right) & 0
\end{array}\right)\ .
\end{equation}

\subsection{Appendix C: Different field configurations in the high frequency regime}

We consider in detail three main configurations for the AC field: 
\begin{enumerate}
\item We fix $A_{x}=A_{y}\sqrt{3}$ and vary $A_{y}$ and $\varphi$ independently.
In this configuration the relation $A_{x}=A_{y}\sqrt{3}$ for the
AC field amplitudes compensates the asymmetry in x-y of the honeycomb
lattice. It gives rise to very simple expressions for the renormalized
hoppings. 
\item We fix $\varphi=\pi/2$ and vary $A_{x}$ and $A_{y}$ independently.
Note that this is reduced to circular polarization if $A_{x}=A_{y}$. 
\item We fix $\varphi=0$, and vary $A_{x}$ and $A_{y}$ independently.
This case contains all possible configurations of linearly polarized
fields. 
\end{enumerate}
The functions $\mathcal{A}_{j}$ and $\Psi_{j}$ for the condition
$A_{y}=A_{x}/\sqrt{3}$ are given by: 
\begin{eqnarray*}
\mathcal{A}_{2,1} & = & \frac{A_{x}a}{\sqrt{2}},\ \mathcal{A}_{3}=A_{x}a\ \\
\Psi_{2,1} & = & \arctan\left(\frac{\sin\left(\varphi\right)}{\left[1\pm\cos\left(\varphi\right)\right]}\right),\ \Psi_{3}=0\ ,
\end{eqnarray*}
with the renormalized hoppings $t_{3}^{F}=tJ_{0}\left(A_{x}a\right)$,
$t_{2}^{F}=tJ_{0}\left(aA_{x}\cos\left(\frac{\varphi}{2}\right)\right)$,
and $t_{1}^{F}=tJ_{0}\left(aA_{x}\sin\left(\frac{\varphi}{2}\right)\right)$.
The figure \ref{fig:Merging-phases} shows the corresponding phase diagram.

\begin{figure}[H]

\begin{centering}
\includegraphics[scale=0.6,angle=-90]{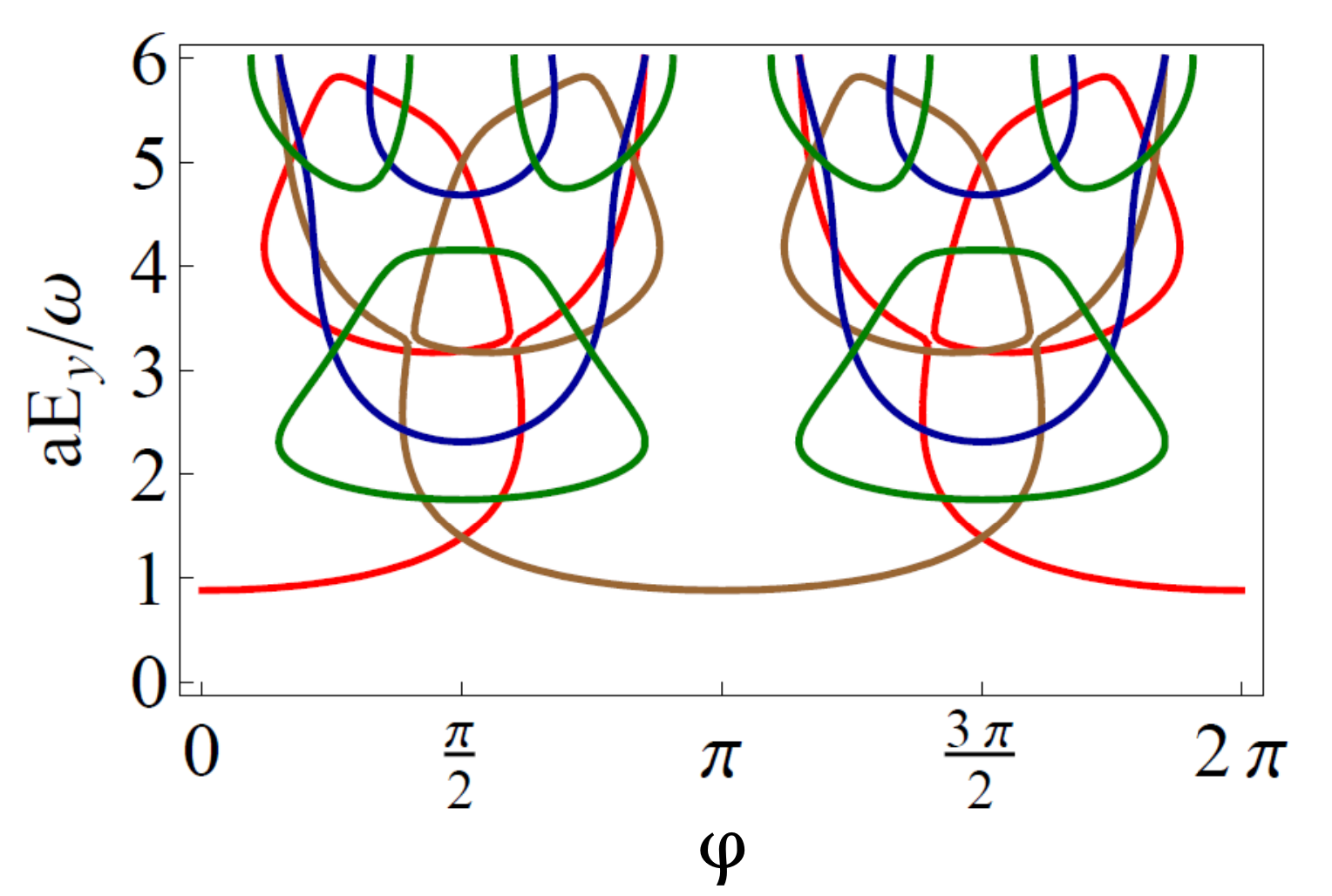} 
\par\end{centering}

\caption{\label{fig:Merging-phases}Phase diagram as a function of the external
AC electric field parameters $\varphi$ and $E_{y}$ for the condition
$A_{x}=\sqrt{3}A_{y}$. The different colors label the four inequivalent
$\text{M}_{j}$ points for the merging.}
\end{figure}

The color code is given by:

\begin{center}
\begin{tabular}{|c|c|}
\hline 
Renormalization  & Merging point\tabularnewline
\hline 
\hline 
$t_{1}^{F}+t_{2}^{F}+t_{3}^{F}=0$  & $\text{M}_{0}$ (Green)\tabularnewline
\hline 
$t_{1}^{F}=t_{2}^{F}+t_{3}^{F}$  & $\text{M}_{1}$ (Brown)\tabularnewline
\hline 
$t_{2}^{F}=t_{1}^{F}+t_{3}^{F}$  & $\text{M}_{2}$ (Red)\tabularnewline
\hline 
$t_{3}^{F}=t_{2}^{F}+t_{1}^{F}$  & $\text{M}_{3}$ (Blue)\tabularnewline
\hline 
\end{tabular}
\par\end{center}

Note that this field configuration allows to merge two or four Dirac
points (PDP) simultaneously (Fig.\ref{fig:Merging-phases}).

For the case of phase difference $\varphi=\pi/2$, the functions are
given by: 
\begin{eqnarray*}
\mathcal{A}_{2,1} & = & \frac{a}{2}\sqrt{A_{x}^{2}+3A_{y}^{2}},\\
\Psi_{2,1} & = & \arctan\left(\frac{\sqrt{3}A_{y}}{A_{x}}\right),\ \mathcal{A}_{3}=A_{x}a,\ \Psi_{3}=0.
\end{eqnarray*}
 with a corresponding phase diagram plotted in Fig.\ref{fig:Merging-phases2}.
Note that the case of circular polarization is obtained by tracing
a line $A_{x}=A_{y}$. Interestingly, the case $\varphi=\pi/2$ only
presents a trivial insulating phase (description in the main text),
and the two merging lines $M_{1}$ and $M_{2}$ overlap. Thus it is
possible to find the merging of one or four PDP, being the latter
related with full localization of the electrons.

\begin{figure}[H]

\begin{centering}
\includegraphics[scale=0.5]{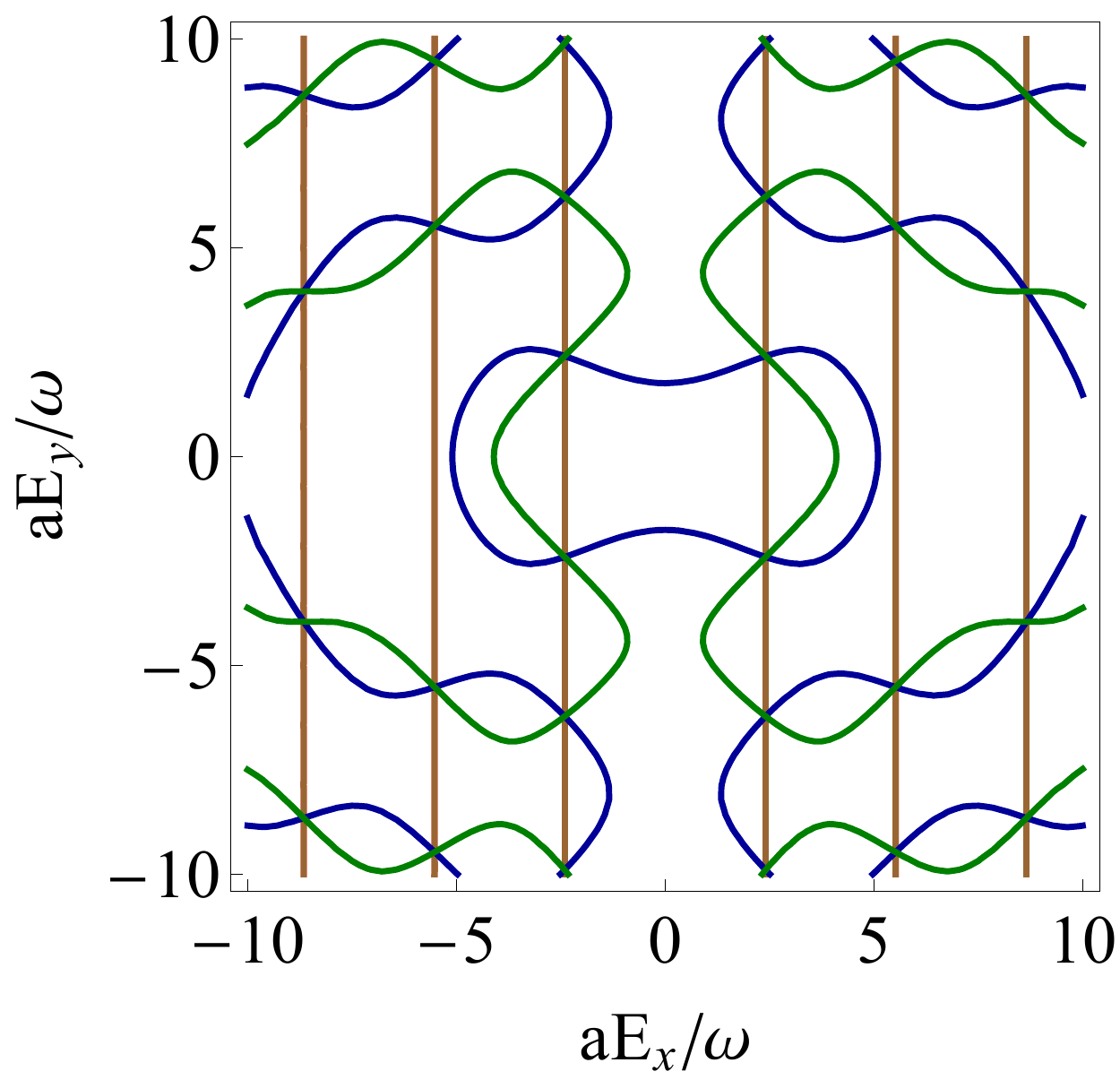} 
\par\end{centering}

\caption{\label{fig:Merging-phases2}Phase diagram as a function of the external
ac electric field parameters $E_{x}$ and $E_{y}$ for the condition
$\varphi=\pi/2$. The different colors label the four inequivalent
$\text{M}_{j}$ points for the merging. Note that in this case the
merging lines $\text{M}_{1}$ and $\text{M}_{2}$ overlap, giving
rise to the existence of two and four simultaneous crossings.}
\end{figure}

For linear polarization ($\varphi=0$) the functions are given by:
\begin{eqnarray*}
\mathcal{A}_{2,1} & = & \frac{a}{2}\sqrt{A_{x}^{2}+3A_{y}^{2}\pm2\sqrt{3}A_{x}A_{y}},\\
\Psi_{2,1} & = & 0,\ \mathcal{A}_{3}=A_{x}a,\ \Psi_{3}=0.
\end{eqnarray*}
 The corresponding phase diagram is plotted in Fig.\ref{fig:Merging-phases3}.
As we can see, this phase diagram has analogies with the one in Fig.\ref{fig:Merging-phases}.

\begin{figure}[H]

\begin{centering}
\includegraphics[scale=0.5]{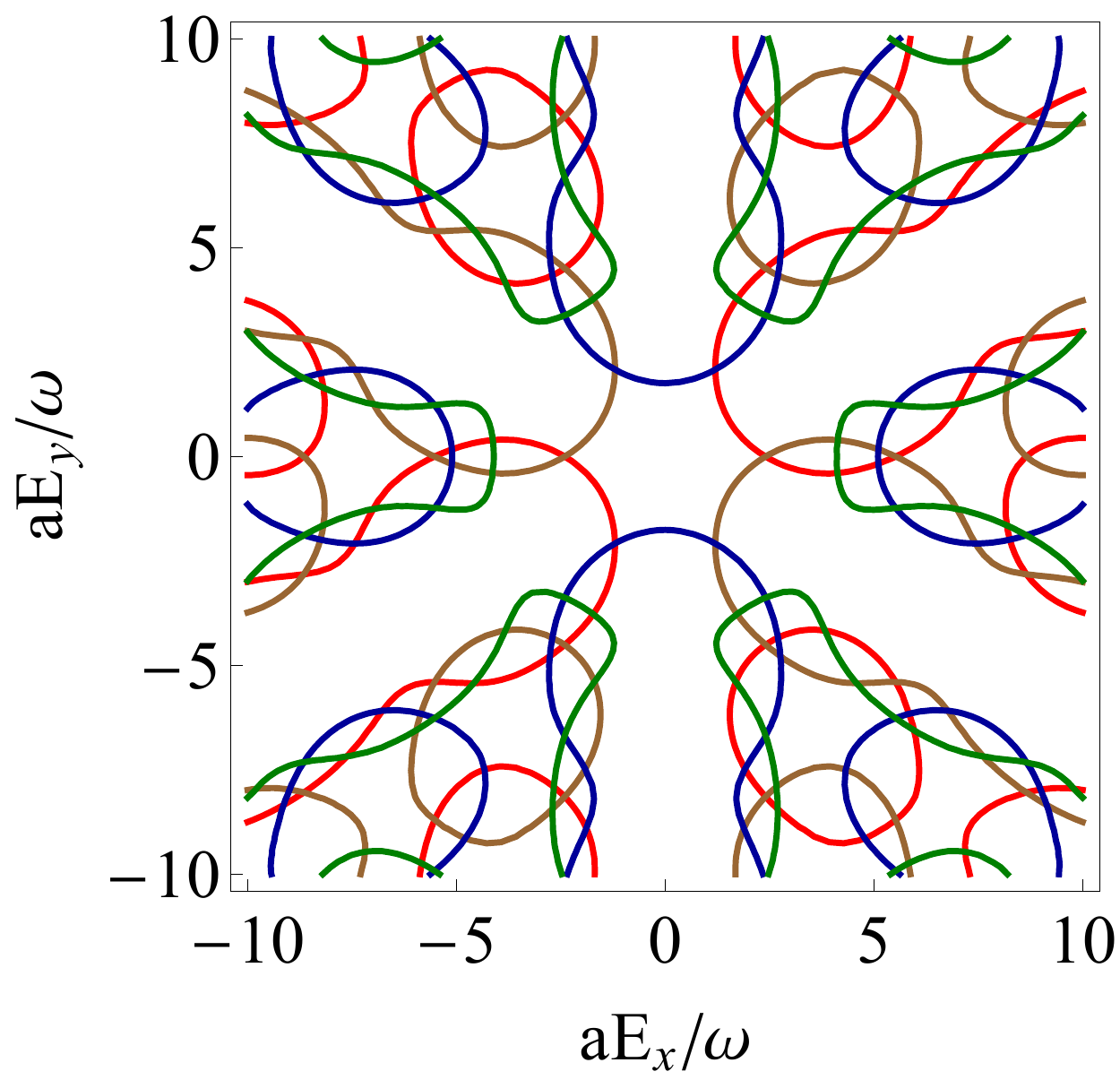} 
\par\end{centering}

\caption{\label{fig:Merging-phases3}Merging lines as a function of the external
ac electric field parameters $E_{x}$ and $E_{y}$ for the condition
$\varphi=0$. The different colors label the four inequivalent $\text{M}_{j}$
points for the merging.}
\end{figure}

In addition, in this plot it is clearly seen the imprint of the lattice
symmetry (hexagonal pattern) around the $\text{SM}_{0}$ phase (the
one at $E_{x}=E_{y}=0$ which is connected to undriven graphene).
This can be expected from the fact that linearly polarized fields
do not modify the phase adquired by the electron during the hopping
($\Psi_{i}=0$ for all $i$). Thus, the effective lattice with renormalized
hoppings must conserve the hexagonal symmetry. This effect can be
understood as well in terms of the hoppings: Tuning the electric field
orientation, we favor one of the three different hoppings. It leads
to three different merging lines (blue, red and brown) and the six-fold
symmetry occurs because of the positive/negative value of the field
amplitude.\end{widetext}
\end{document}